\documentclass[letter,english]{article}
\usepackage{emulateapj,onecolfloat,graphics}

\begin{document}

\twocolumn
[
\title{The halo model and numerical simulations}
\author{Martin White, Lars Hernquist, Volker Springel}
\affil{Harvard-Smithsonian Center for Astrophysics, Cambridge, MA 02138}

\begin{abstract}
\noindent
Recently there has been a lot of attention focussed on a virialized halo-based
approach to understanding the properties of the matter and galaxy power
spectrum.  A key ingredient in this model is the number and distribution of
galaxies within dark matter halos as a function of mass.
This quantity has been predicted from semi-analytic modeling and from fits
to observational data.
Here we present predictions for the occupation number and spatial distribution
of sub-halos based on a high-resolution hydrodynamical simulation including
cooling, star-formation and feedback.
\end{abstract}
 ]

\section{Introduction}

For hierarchical models of structure formation based on the inflationary
cold dark matter paradigm, the clustering and evolution of dark matter halos
is now quite well understood and can be reliably simulated using N-body
techniques.
In contrast, a theory for the formation and evolution of galaxies is one
of the central unsolved problems in cosmology.
Such a theory should at the very least specify the relationship between
galaxies and the dark matter halos in which they reside, and it is this
question which we wish to address.

Several authors
(Ma \& Fry~\cite{MaFry1,MaFry2}; Seljak~\cite{Sel};
 Peacock~\cite{Pea}; Peacock \& Smith~\cite{PeaSmi})
have recently developed a new way of describing the non-linear clustering of
dark matter and galaxies.
This formalism postulates that all the mass in the universe lies 
in halos of various masses 
(Press \& Schechter~\cite{PreSch}) and further that on large scales
the halos cluster according to linear theory while on small scales the power
is dominated by halo profiles 
(Neyman, Scott \& Shane~\cite{NeyScoSha}; Peebles~\cite{Pee}).
The galaxies are assigned to halos using simple rules.
While this model requires many ingredients to be fixed by numerical
experiments (typically N-body simulations or semi-analytic modeling)
it provides a useful structure for thinking about gravitational clustering
which gives insights into several outstanding problems
(see e.g.~Seljak~\cite{Sel};
 Peacock \& Smith~\cite{PeaSmi};
 Seljak, Burwell \& Pen~\cite{SelBurPen};
 Atrio-Barandela \& Mucket~\cite{AtrMuc};
 Cooray, Hu \& Miralda-Escude~\cite{CorHuMir};
 White~\cite{Whi};
 Scoccimarro et al.~\cite{SSHJ};
 Guzik \& Seljak~\cite{GuzSel};
 Seljak~\cite{SelB};
 Sheth et al.~\cite{SDHS}).

Here we shall be primarily interested in the 2-point function of galaxy
clustering, either in Fourier space (the power spectrum) or real space
(the correlation function).
The 2-point function is one of the most fundamental quantities in
large-scale structure.  It is robust, but sensitive to several cosmological
parameters such as the Hubble constant, the matter density and of course the
primordial power spectrum.
The key insight of the halo model is that an accurate prediction of this
quantity requires a knowledge of the occupation number of galaxies in dark
matter halos and their spatial distribution
(see also Benson et al.~\cite{BCFBL}).
With only these ingredients the model can make predictions about a wide
variety of quantities of interest.
In addition, these quantities allow one to ``graft'' galaxies onto pure
N-body simulations of large-scale structure to make mock galaxy catalogues
(Benson et al.~\cite{BCFBL}; Peacock \& Smith~\cite{PeaSmi}).

The occupation number distribution, $\langle N\rangle(M)$ and
$\langle N(N-1)\rangle(M)$ at the 2-point level, and the spatial distribution
of galaxies thus make an ideal point of contact between observations and
theory.  They can, in principle, be predicted from a model of galaxy formation,
and they can be measured observationally.

Seljak~(\cite{Sel}) measured $\langle N\rangle(M)$ from semi-analytic
models of structure formation (Kauffmann et al.~\cite{KCDW}; but see
also Somerville \& Primack~\cite{SomPri} and Benson et al.~\cite{BCFBL})
and used this to make predictions for the clustering of galaxies.
A similar procedure has been followed by
Wechsler et al. (2000)
where simulations are used in place of analytic techniques.
Peacock \& Smith~(\cite{PeaSmi}) and Scoccimarro et al.~(\cite{SSHJ}) took
the opposite approach of trying to infer this distribution from observations
as a check on semi-analytic modeling.
The agreement between the two methods is at most qualitative at present.
Here we present predictions for these functions from a cosmological
hydrodynamic simulation which includes star-formation and feedback.

\section{The simulation}

Throughout, we shall use a new simulation of the Ostriker \& Steinhardt
(\cite{OstSte}) concordance model, which has
$\Omega_{\rm m}=0.3$, $\Omega_\Lambda=0.7$,
$H_0=100\,h\,{\rm km}{\rm s}^{-1}{\rm Mpc}^{-1}$ with $h=0.67$,
$\Omega_{\rm B}=0.04$, $n=1$ and $\sigma_8=0.9$
(corresponding to $\delta_H=5.02\times 10^{-5}$).
This model yields a reasonable fit to the current suite of cosmological
constraints and as such provides a good framework for making realistic
predictions.

We have used the {\sc Tree/SPH\/} code {\sc Gadget\/}
(Springel, Yoshida \& White~\cite{SprYosWhi}) to run a
$2\times 300^3=54$ million particle simulation of this model in a
periodic box of size $33.5\,h^{-1}$Mpc.  Equal numbers of gas and dark
matter particles were employed, so $m_{\rm dark}=1\times 10^8\,h^{-1}M_\odot$
and $m_{\rm gas}=1.5\times 10^7\,h^{-1}M_\odot$ and the gravitational
softening was $6\,h^{-1}$kpc, fixed in comoving coordinates.
The simulation was started at redshift $z=99$ and evolved to $z=1$.

In addition to the gravitational interactions and adiabatic hydrodynamics,
the code follows radiative cooling and heating processes in the presence
of a UV radiation field in essentially the same way as described in
Katz, Weinberg \& Hernquist~(\cite{KatWeiHer}).
We model the UV radiation field using a Haardt \& Madau~(\cite{HaaMad})
spectrum with reionization at $z=6$.
Star formation (and feedback) is handled using a modification of the
``multi-phase'' model of Yepes et al.~(\cite{YepKatKhoKly}) and
Hultman \& Pharasyn~(\cite{HulPha}).
Each SPH particle is assumed to model a co-spatial fluid of ambient hot
gas, condensed cold clouds, and stars.  Hydrodynamics is only followed for
the hot gas phase, but the cold gas and stars are subject to gravity,
add inertia, and participate in mass and energy exchange processes with the
ambient gas phase.
The algorithm will be described in more detail in a forthcoming paper.

{}From the simulation outputs at $z=1$ and 3 we have constructed catalogues
of halos and their sub-halos using the algorithm {\sc Subfind\/} described
in detail in Springel et al.~(\cite{SWTK}).  First, the Friends-of-Friends
algorithm (with a linking length of $0.15\bar{n}^{-1/3}$) is used to define
a parent halo catalogue, and then bound sub-halos within each parent are
identified.  These sub-halos typically consist of cold, dense gas that has
been able to efficiently cool and form stars and should be identified with
galaxies in the real universe, at least statistically.

We have used a linking length of 0.15 rather than the more canonical 0.2 in
defining the parent halos, because we found for the larger linking length
neighboring halos were being linked with one then identified as a sub-halo
of the other.
While this problem is not eliminated entirely using 0.15, it is
significantly reduced.  Only very close halo pairs or triplets, possibly
in the process of merging, are joined with this linking length.
In some instances the FOF algorithm finds halos which are not bound; these
halos are not included in the analysis.

At $z=1$ there are a total of 31,737 halos and 34,529 sub-halos with an
average star formation rate of $0.35 M_\odot\, {\rm yr}^{-1}$.
At $z=3$ the numbers are 37,940 and 39,579 respectively, with an average
star formation rate of $0.25 M_\odot\, {\rm yr}^{-1}$.

\section{Halo occupation number}

\begin{figure}
\begin{center}
\resizebox{3.5in}{!}{\includegraphics{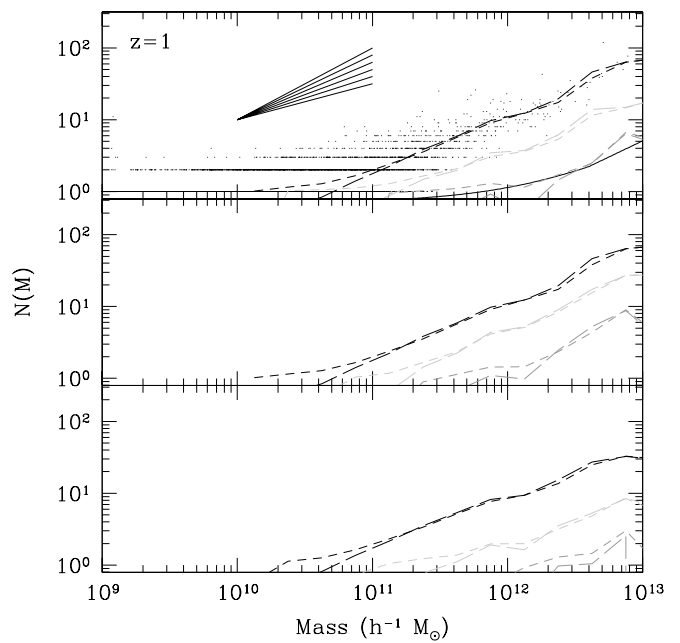}}
\end{center}
\caption{The number of sub-halos per FOF group as a function of parent halo
mass in the simulation at $z=1$.
(Top panel) There is 1 dot per halo.  Also shown are $\langle N\rangle$
(long dashed) and $\sqrt{\langle N(N-1)\rangle}$ (short dashed) as a function
of mass for:
(upper lines) all sub-halos,
(middle) sub-halos with $M>10^{10}\,h^{-1}M_\odot$ and
(lower) sub-halos with $M>10^{11}\,h^{-1}M_\odot$.
Note that the number of sub-halos per parent is close to Poisson when
$\langle N\rangle\gg 1$ but sub-Poisson for small $\langle N\rangle$,
the mean is well approximated by a power-law and it increases more slowly
than $M$.
The ray of solid lines have slopes from $0.5$ to $1$ for comparison.
The lower solid line is a prediction from semi-analytic modeling (see text).
(Middle panel)As above but with cuts on stellar mass:
(upper lines) all sub-halos,
(middle) sub-halos with $M_{\rm star}>10^{9}\,h^{-1}M_\odot$ and
(lower) sub-halos with $M_{\rm star}>10^{10}\,h^{-1}M_\odot$.
(Bottom panel) As above but with cuts on star-formation rate:
(upper lines) all sub-halos,
(middle) sub-halos with SFR $>1M_\odot/$yr,
(lower) sub-halos with SFR $>10M_\odot$/yr.}
\label{fig:nofm_z1}
\end{figure}

\begin{figure}
\begin{center}
\resizebox{3.5in}{!}{\includegraphics{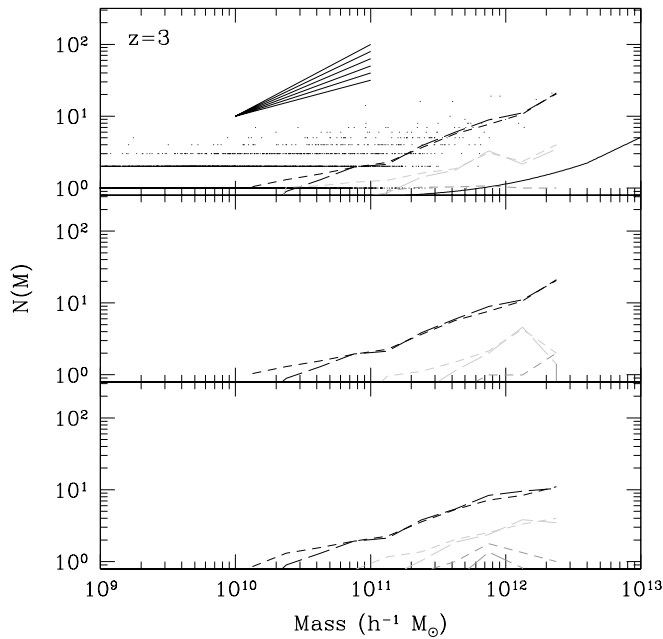}}
\end{center}
\caption{As for Fig.~\protect\ref{fig:nofm_z1} but at $z=3$.}
\label{fig:nofm_z3}
\end{figure}

We show in Figs.~\ref{fig:nofm_z1}, \ref{fig:nofm_z3} the distribution
$N_{\rm sub}(M)$ from the simulation.
Ideally we would identify the sub-halos as galaxies by some observational
property, such as luminosity or color.
Unfortunately we are unable to reliably compute such properties from this
simulation.
Even though we track the star-formation rate in the simulation, the 
outputs are
too infrequent to graft on population synthesis codes, and any transformation
we apply to the outputs would simply introduce additional uncertainties
and modeling into our predictions.
Thus we have focussed on the number of sub-halos as a function of sub-halo
(total) mass, sub-halo stellar mass and sub-halo ``instantaneous''
star-formation rate which we can reliably extract from the simulation.
Additional modeling will be presented elsewhere.

Since the FOF groups can be irregular and their mass difficult to interpret,
we compute spherically averaged masses using the group particles.
The center of the halo is taken to be the point of minimum potential,
considering the halo particles in isolation.  This corresponds very closely
to the densest particle and the most bound particle for all but the most
disturbed systems.
For the total halo mass we use the mass interior to a radius ($r_{500}$)
inside of which the mean density is 500 times the critical density.
For some sub-halos the halo is truncated before this radius in which case
we use all of the bound mass in the halo.
The stellar mass is simply the mass in stars within $r_{500}$ and the star
formation rate is also summed over all star-forming particles within this
radius.

Note that for parent halos of a fixed mass there is a distribution of
sub-halo occupation numbers.  Statistically the distribution seems to
be quite close to Poisson for $\langle N\rangle\gg 1$ but has less scatter
than the Poisson prediction for $\langle N\rangle\sim 1$, as is also seen
in semi-analytic models.
By definition all bound halos host at least one sub-halo.
For all panels in Fig.~\ref{fig:nofm_z1} the distribution $\langle N\rangle$
is close to a power-law, with a slope of $\sim 0.8-0.9$, i.e.~the number of
sub-halos increases more slowly than the mass.
At $z=3$ we again find that $\langle N\rangle$ is well fit by a power-law,
with a slightly shallower slope than at $z=1$, being closer to $0.7$, again
increasing more slowly than the mass.

One could imagine that the decrease in $\langle N\rangle/M$ may arise because
the larger halos have longer gas cooling times, because mergers are more
efficient or because a larger fraction of the total mass lies ``outside'' of
the sub-halos.
In our simulation we find that the both the mean sub-halo mass (as a fraction
of the parent halo mass) and the fraction of the halo mass in sub-halos
decline with increasing parent halo mass.  At the low mass end the mean
sub-halo mass and the fraction of the mass in sub-halos, using $M_{500}$,
are near 30\%.  For the higher mass halos this declines to $<1\%$ and $10\%$
respectively.
This suggests that all three effects are operating to some degree.

The predictions of the semi-analytic models of Kauffmann et al.~(\cite{KCDW}),
as fit by Sheth \& Diaferio~(\cite{SheDia}), are reproduced in
Figs.~\ref{fig:nofm_z1}, \ref{fig:nofm_z3} as the lower solid line extending
off the right of the figure.
We can see in Fig.~\ref{fig:nofm_z1} that the semi-analytic models curve
roughly agrees with our numerical results for sub-halos more massive than
$10^{11}M_\odot$, but has fewer ``galaxies'' than we have sub-halos if we
include less massive objects.
This could partly be due to limited mass resolution in the simulations onto
which are grafted the semi-analytic recipes, but is also because the
semi-analytic formalism has explicitly inefficient star-formation in low mass
halos.

\section{Sub-halo spatial distribution}

Another ingredient in the halo model is that all parent halos have
NFW (Navarro, Frenk \& White~\cite{NFW}) profiles\footnote{Or revised
profiles which rise more steeply than $r^{-1}$ as $r\to 0$.}, and that
the satellite galaxy distribution follows the dark matter distribution.
This obviously implies that different galaxy types trace each other within a
given halo, which has not been shown.
It remains unclear whether any type of galaxy follows the mass
distribution.

\begin{figure}
\begin{center}
\resizebox{3.5in}{!}{\includegraphics{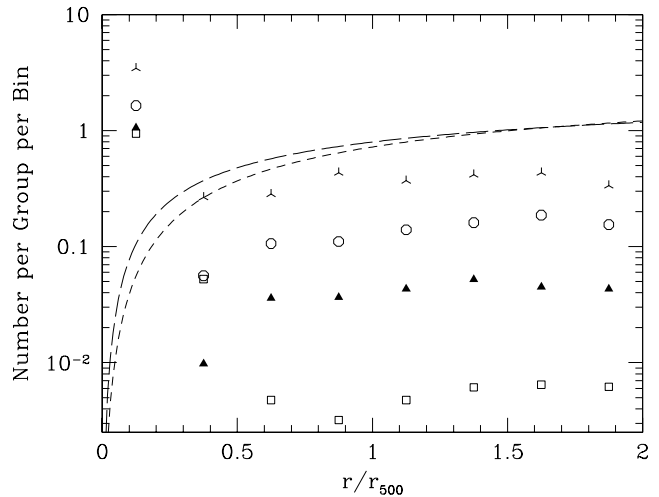}}
\end{center}
\caption{The number of sub-halos per FOF halo in the simulation at $z=1.0$
as a function of radius from the center of the parent halo.
All radii are scaled to the $r_{500}$ of the parent halo.
We have binned the parent halos by $\log_{10}(M/h^{-1}M_\odot)$:
(open squares): $10-10.5$, (solid triangles) $10.5-11$, (open circles)
$11-11.5$, (stars) $11.5-12$.
The dashed lines are cumulative mass within $r$ for two NFW profiles with
$c=5$ and $c=10$, normalized to unity at $r_{200}$.}
\label{fig:radial}
\end{figure}

To investigate this question we have calculated the mean number of sub-halos
in radial bins, scaled to $r_{500}$ of the parent halo.
Fig.~\ref{fig:radial} shows that there is an enhanced probability for
sub-halos to reside at the center of the parent halo (i.e.~one galaxy always
resides at the center of the dark matter halo) and then the probability is
rather flat with radial distance from the center.
Because FOF with a linking length of 0.15 can occasionally link neighboring
structures, a few sub-halos are found beyond the virial radius.

For comparison we show the cumulative mass within $r$ for two NFW profiles,
one with $c\equiv r_{200}/r_s=5$ and one with $c=10$.  The total mass is
(arbitrarily) normalized to unity at $r_{200}\sim 2r_{500}$.  Apart from
the central enhancement the sub-halos do approximately seem to follow the
NFW profile, although they seem to be slightly more centrally concentrated
than NFW would predict.

Very few of our halos have enough sub-halos for a robust determination of
the velocity distribution of the sub-halos.  However since the dark matter
and the sub-halos live in the same potential and have similar spatial
distributions, we expect that any degree of ``velocity bias'' would be small.

\section{Conclusions}

Many of the features of the power spectrum of density fluctuations in the
universe can be simply understood in a model based on virialized halos.
Important ingredients in the model are that the halos be biased tracers of
the linear power spectrum and have a uniform profile with a correlation
between the internal structure and the mass which should span a wide range
sampling a Press-Schechter (\cite{PreSch}) like mass function.

In order to understand the clustering of galaxies within such a model it is
important to have a prediction for the occupation number distribution of
galaxies in a halo of a certain mass and their spatial distribution within
the halo.  This quantity can be predicted by semi-analytic models of
galaxy formation or estimated from observations.  Here we have presented
an estimate of this function spanning a wide range in halo masses from
hydrodynamical simulations including cooling, star formation and feedback.
(Previous efforts to study galaxy clustering directly from hydrodynamical
simulations
[e.g. Katz et al.~\cite{KHW92,KHW99}; Gardner et al.~\cite{GKHW97,GKHW99};
Dav\'e et al.~\cite{DHKW}; Pearce et al.~\cite{Pearce99}] typically had
insufficient dynamic range to  estimate the galaxy occupation number over
such a wide range in halo mass.)

We find that the number of galaxies per halo is very close to a Poisson
distribution, with a mean which grows more slowly than the mass of the
parent halo.  This holds true regardless of whether we select sub-halos
based on total mass, stellar mass or star formation rate.
The spatial distribution of sub-halos within their parent halos appears to
follow the distribution of the total mass relatively well, with the number
of sub-halos per spherical shell being approximately constant at large
radius.  Closer to the center of the halo there are more sub-halos than
a halos-follow-mass model would predict.

\section*{Acknowledgments}

We thank C.-P. Ma, J.~Peacock and U.~Seljak for helpful conversations about
the halo model and the $\langle N\rangle(M)$ distribution.  This work was
supported in part by the Alfred P. Sloan Foundation and the National
Science Foundation, through grants PHY-0096151, ACI96-19019 and AST-9803137.

\end{document}